# LIGHT-SABRE hyperpolarizes 1-$^{13}$C-pyruvate continuously, without magnetic field cycling

*Andrey N. Pravdivtsev[a], Kai Buckenmaier[b]\*, Nicolas Kempf[b], Gabrielle Stevanato[c],[d], Klaus Scheffler[b],[e], Joern Engelmann[b], Markus Plaumann[f], Rainer Koerber[g], Jan-Bernd Hövener[a], and Thomas Theis[b],[h],[i]\**

[a] Section Biomedical Imaging, Molecular Imaging North Competence Center (MOIN CC), Department of Radiology and Neuroradiology, University Medical Center Kiel, Kiel University, Am Botanischen Garten 14, 24118, Kiel, Germany, E-mail: andrey.pravdivtsev@rad.uni-kiel.de, jan.hoevener@rad.uni-kiel.de

[b] High-Field Magnetic Resonance Center, Max Planck Institute for Biological Cybernetics, Max-Planck-Ring 11, 72076, Tübingen, Germany E-mail: kai.buckenmaier@tuebingen.mpg.de

[c] Department of Chemical Sciences, University of Padova, Via Marzolo 1, 35131 Padova, Italy

[d] NMR Signal Enhancement Group, Max Planck Institute for Multidisciplinary Sciences, Am Fassberg 11, 37077 Göttingen, Germany.

[e] Department for Biomedical Magnetic Resonance, University of Tübingen, Tübingen, Germany

[f] Otto-von-Guericke University, Medical Faculty, Institute of Biometry and Medical Informatics, Leipziger Str. 44, 39120 Magdeburg, Germany

[g] Physikalisch-Technische Bundesanstalt, Department 'Biosignals', Abbestraße 2-12, 10587 Berlin, Germany

[h] Departments of Chemistry and Physics, North Carolina State University, 27695, Raleigh, NC, USA E-mail: ttheis@ncsu.edu

[i] Joint UNC-NC State Department of Biomedical Engineering, North Carolina State University, 27606, Raleigh NC, USA

AUTHOR INFORMATION

**Corresponding Author**

* Kai.Buckenmaier@tuebingen.mpg.de, ttheis@ncsu.edu

ABSTRACT. Nuclear spin hyperpolarization enables real-time observation of metabolism and intermolecular interactions *in vivo*. 1-$^{13}$C-Pyruvate is the leading hyperpolarized tracer currently under evaluation in several clinical trials as a promising molecular imaging agent. Still, the quest for a simple, fast, and efficient hyperpolarization technique is ongoing. Here, we describe that continuous, weak irradiation in the audio-frequency range of the $^{13}$C spin at 121 µT magnetic field (~twice Earth's field) enables spin order transfer from parahydrogen to $^{13}$C magnetization of 1-$^{13}$C-pyruvate. These so-called LIGHT-SABRE pulses couple nuclear spin states of parahydrogen and pyruvate via the *J*-coupling network of reversibly exchanging Ir-complexes. Using ~100% parahydrogen at ambient pressure, we polarized 51 mM of 1-$^{13}$C-pyruvate in the presence of 5.1 mM Ir-complex continuously and repeatedly to a polarization of 1.1% averaged over free and catalyst-bound pyruvate. The experiments were conducted at -8°C), where almost exclusively bound pyruvate was observed, corresponding to an estimated 11% polarization on bound pyruvate.

The obtained hyperpolarization levels closely match those obtained via SABRE-SHEATH under otherwise identical conditions. The creation of three different types of spin orders was observed: transverse $^{13}$C magnetization along the applied magnetic field, $^{13}$C $z$-magnetization along the main field $B_0$, and $^{13}$C-$^1$H $zz$-spin-order. With a superconducting quantum interference device (SQUID) for detection, we found that the generated spin orders result from tiny $^1$H-$^{13}$C $J$-coupling interactions, which are not visible even with our narrow linewidth below 0.3 Hz.

**TOC GRAPHICS**

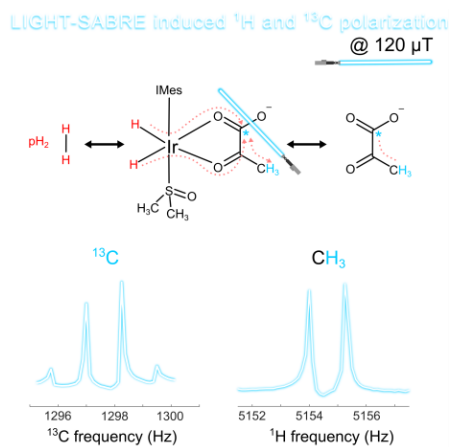

Nuclear spin hyperpolarization of 1-$^{13}$C-pyruvate is a successful example of translating quantum technology into clinical practice.[1–3] Currently, dissolution dynamic nuclear polarization (dDNP) is the leading method to hyperpolarize pyruvate[4–7] and other small-molecule metabolites[8–10]. An alternative approach to dDNP is parahydrogen-induced polarization (PHIP), which is faster and simpler, however, less developed.[11,12] Hydrogenative PHIP has been used to produce close to unity hyperpolarization of pyruvate esters at high concentrations,[13–16] which was facilitated by the development of perdeuterated and $^{13}$C labeled pyruvate esters.[15–18] A non-hydrogenative PHIP variant is signal amplification by reversible exchange (SABRE),[19] which allows continuous (re-)hyperpolarization of selected molecules. SABRE can also be used to hyperpolarize pyruvate and does not require hydrogenation. Instead, transient interactions with an Ir-complex are sufficient to hyperpolarize pyruvate without chemical modification, as depicted in **fig. 1a**.[20–23] The polarization transfer from parahydrogen (pH$_2$) to pyruvate was demonstrated at ultralow magnetic fields below 1 µT, where the pH$_2$-derived protons and the $^{13}$C nucleus of pyruvate become strongly coupled and anticrossings of nuclear spin energy levels occur. This principle is used in the technique known as SABRE in shield enables alignment transfer to heteronuclei, SABRE-SHEATH.[24–28] After hyperpolarization in SABRE-SHEATH experiments, the magnetic field is typically elevated to a few Tesla for detection. An alternative approach to transfer the pH$_2$ spin order to $^1$H, $^{13}$C, or $^{15}$N is to use RF pulse sequences at a constant field,[29–35] however, this approach has not yet been demonstrated for pyruvate.

Here, we demonstrate continuous hyperpolarization of 1-$^{13}$C-pyruvate using RF pulses at a magnetic field of 121 µT (~ twice Earth's field). The gained hyperpolarization levels are as large as those obtained with SABRE-SHEATH under otherwise identical conditions: -8°C, 51 mM pyruvate, 5.1 mM Ir-complex in MeOH, ~100% pH$_2$ at ambient pressure. The observed hyperpolarization is on the order of ~1.1% averaged over free and bound species. Note that at -8°C, only bound pyruvate was observed, and polarization was estimated to be ~11%.[22] We also demonstrate that three different types of spin orders can be generated: transverse ($x$), longitudinal ($z$), and two-spin order ($zz$). First, transverse ($x$) polarization can be generated along an applied, on-resonance RF field. This was the most efficient mechanism for polarizing the $^{13}$C nucleus of 1-$^{13}$C-pyruvate studied here. Second, we demonstrate that $z$-magnetization can be generated by applying a weak RF field slightly off-resonance. This mechanism is somewhat less efficient than on-resonance irradiation. We also describe the creation of $zz$-order on $^{13}$C-$^{1}$H spin pairs in 1-$^{13}$C-pyruvate for the on-resonance case.

Since the experiments were conducted at relatively low temperatures of -8°C, pyruvate exchange is strongly suppressed.[21,22] As in previous work, here we also observe that hydrogen exchange, on the other hand, is still highly efficient. **Figure 1b-c** illustrates possible exchange mechanisms that allow for hydrogen exchange without pyruvate exchange under the assumption that hydrogen exchange is an associative mechanism, as supported by previous work.[36,37] We hypothesize that the pyruvate-SABRE system enables efficient hydrogen exchange because the pyruvate binds in a bidentate fashion, which allows for hydrogen exchange via partial dissociation of pyruvate (**fig. 1b**), opening a binding site for parahydrogen without pyruvate exchange all the way into its free form. A less likely alternative is depicted in **fig. 1c**, where DMSO dissociation allows for H$_2$ association and exchange.

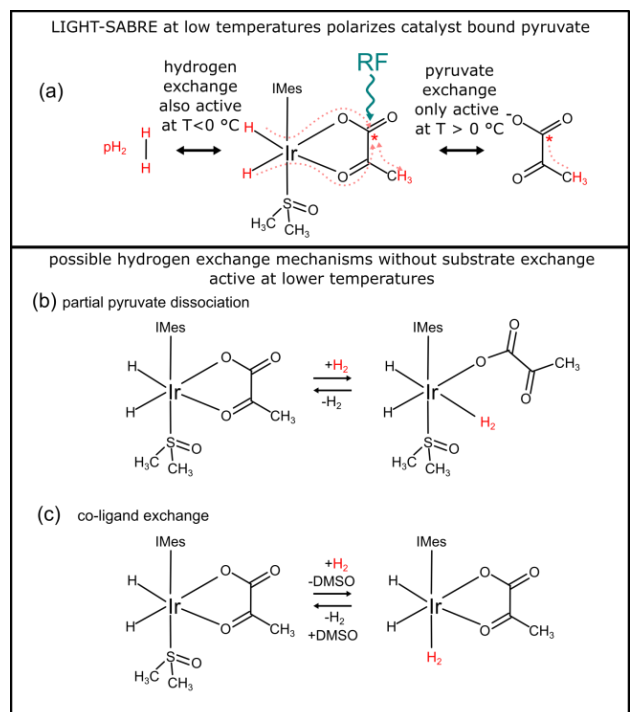

**FIGURE 1.** LIGHT-SABRE scheme of the SABRE-reaction to polarize pyruvate, where chemical exchange and spin order transfer occur on the Iridium-based polarization transfer catalyst. The LIGHT-SABRE RF pulse is applied close to the Larmor frequency of the $^{13}C$ nuclear spin. Hydrogen exchange remains active at lower temperatures, whereas pyruvate exchange is only active at elevated temperatures. In the present work, all experiments were performed at temperatures below 0°C. Panels b and c present possible exchange mechanisms that allow for hydrogen exchange without substrate exchange. The partial pyruvate dissociation mechanism shown in b) seems more plausible than the co-ligand exchange mechanism shown in c, while it cannot be ruled out at this time. Here, IMes is defined as 1,3-bis(2,4,6-trimethylphenyl)imidazol-2-ylidene; DMSO is added as a co-substrate to optimize exchange rates; red dotted arrows indicate polarization transfer between nuclei.

## RESULTS and DISCUSSION

In the current demonstration, 1-$^{13}C$-pyruvate SABRE polarization was generated and observed at a constant magnetic field of $B_0 = 121$ µT. The Faraday coil benchtop NMR spectrometers at such low field strengths are not sensitive, so we opted for using a superconducting quantum interference device (SQUID) based ultra-low field spectrometer.[38–40] Since SQUIDs are broadband detectors, signals from $^1$H and $^{13}$C can be detected simultaneously. The noise level of the employed SQUID-detector is on the order of 1 fT Hz$^{-½}$, resulting in signal-to-noise ratios of above 8,000 in a single shot when detecting the hyperpolarized 1-$^{13}$C pyruvate. The SQUID NMR system sits inside a three-layered shielding chamber for DC and RF fields with a residual magnetic field below 10 nT at its center. The capability of field cycling makes it a versatile system for all kinds of *in vitro* SABRE experiments.

Out of the existing spin order transfer sequences, low-irradiation generation of high tesla – SABRE (LIGHT-SABRE)[33] is probably the simplest because it only requires a single, weak and constant irradiation close to the Larmor frequency of the targeted nucleus, also referred to as spin-lock induced crossing (SLIC) pulse.[41]

The main idea behind SLIC and LIGHT-SABRE is that a correctly set continuous wave (CW) RF pulse matches two energy levels of the hydride-substrate spin system. Consequently, the spin alignment of pH$_2$ evolves into the polarization of the target spins in the substrate. In contrast, SABRE-SHEATH uses $B_0$ to achieve the same polarization transfer effect, adjusting and matching the energy levels by setting $B_0$. The advantage of using $B_1$, instead of $B_0$ for spin order transfer is that field cycling becomes obsolete, and hyperpolarization could be produced at any magnetic field. In the specific case of 1-$^{13}$C-pyruvate SABRE, the *J*-coupling network is dominated by the hydride-hydride coupling $J_{HH}$ of ~-10.5 Hz[23] such that irradiation with a LIGHT-SABRE pulse on-resonance with the $^{13}$C Larmor frequency and amplitude, $\nu_{CW}^{A}$, set to ~10.5 Hz is expected to give maximum hyperpolarization transfer to the $^{13}$C spin.

When LIGHT-SABRE or related methods are used at much higher, multi-Tesla magnetic fields, then pH$_2$ bubbling must be interrupted during LIGHT-SABRE irradiation to avoid $B_0$ and $B_1$ inhomogeneities. At the ultralow magnetic fields used here, $B_0$ = 121 µT, pH$_2$ can be bubbled through the solution during irradiation and acquisition of the free induction decay (FID) without any noticeable decrease of $B_0$ or $B_1$ field homogeneity because (a) susceptibility effects scale with the magnetic field strength and (b) the audio frequency irradiation can easily be performed with large and homogeneous Helmholtz coils. In our experiments, the full width at half maximum of $^{13}$C lines measured was below 0.3 Hz.

The low magnetic field is also beneficial for 1-$^{13}$C-pyruvate LIGHT-SABRE because it preserves the singlet spin state of pH$_2$ after association with the Ir-complex. Although LIGHT-SABRE was proposed and used at high magnetic fields to polarize pyridine,[33] nicotinamide,[42] 4-amido pyridine,[34] in such cases, pH$_2$ derived protons (IrHH) had the same chemical shifts. This is not the case for pyruvate, where the hydride-hydride chemical shift difference is 2 ppm[21]. To compensate for this problem, one typically has to use strong RF pulses to lock the protons in a singlet state,[43] which will also alter the LIGHT-SABRE conditions.[44] At the low magnetic fields employed here, the hydride-hydride chemical shift difference is much smaller than their mutual J-coupling interaction of $J_{HH}$ ~ -10.5 Hz $>>\Delta\nu_{HH}$, which is the requirement for maintaining a singlet state without spin locking.

In the presented work, LIGHT-SABRE produces a range of different spin orders in the polarized substrate (1-$^{13}$C-pyruvate), including (1) transverse polarization ($S_x$ or $S_y$), which is generated parallel to the LIGHT-SABRE CW pulse in the rotating frame, when applying the LIGHT-SABRE pulse on resonance, (2) longitudinal ($S_z$) or z-polarization of the $^{13}$C nucleus (S), which results when applying LIGHT-SABRE slightly off-resonance, and (3), note that the presence of additional spins (CH$_3$ group) (I) can result in additional polarization of ($^{13}$C-$^{1}$H)-zz two-spin order ($S_z \cdot I_z$) or zz-polarization. To detect and distinguish these three different spin orders, $^{13}$C x-polarization (**fig. 2,3**), $^{13}$C *z*-polarization (**fig. 4**), and ($^{13}$C-$^{1}$H) *zz*-polarization (**fig. 5**), different acquisition schemes were designed and tested. The individual pulse sequences are fully described in "Experimental and Computational Details" section (**fig. 6**).

An iteration between experiments and simulations found the optimal RF conditions for the LIGHT-SABRE experiment. Finally, the simulations were fit to the experimental data showing good agreement as discussed below. The MOIN-spin library[45] was used to simulate all SABRE experiments. All scripts are available in the supporting materials.

First, the generation of transverse (*x*) $^{13}$C magnetization was investigated as a function of the $B_1$ amplitude $\nu_{CW}^{A}$ and the frequency offset from the $^{13}$C Larmor precession frequency $\Delta\nu_{CW}^{frq}$ of the CW pulse with a fixed hyperpolarization time $t_{hyp}$ = 10 s at $B_0$ = 121 µT (**fig. 2**). Here, $t_{hyp}$ = 10 s implies a 10 s long LIGHT-SABRE pulse. We note that pH$_2$ constantly bubbled through the solution, even during acquisition. Under these conditions, the hyperpolarized transverse magnetization showed the highest enhancement at $\Delta\nu_{CW}^{frq}$ = 0 Hz and $\nu_{CW}^{A} \cong 1.1$ µT. This amplitude of the CW pulse corresponds to a $^{13}$C $B_1$ Larmor frequency of 11.8 Hz, which is close to the $J_{HH}$ = -10.5 Hz coupling as theoretically predicted.

The simulations indicated that the width of the LIGHT-SABRE-polarization in the $\nu_{CW}^{A}$ dimension is mainly dependent on the lifetime of the active Ir-complex $\tau_{Ir}$. For the experiments performed at -8°C, the best fitting was achieved with $\tau_{Ir}$ = 31 ms, where this lifetime is primarily dominated by hydrogen exchange at low temperatures. The acquired spectra did not allow us to distinguish between free and Ir-bound 1-$^{13}$C-pyruvate because the chemical shift difference for pyruvate in bulk and coordinated to Ir is about 1.3 ppm, which corresponds to ~1 mHz at $B_0$ = 121 µT, which is, below the FWHM of the corresponding $^{13}$C lines. As described previously,[22] no significant hyperpolarization transfer to free pyruvate occurs at these low temperatures. Therefore, in the present work, the observed hyperpolarization is primarily on the catalyst-bound pyruvate.

In previous SABRE research, the SABRE activity was stopped by the addition of bidentate ligands 2,2-bipyridine or 1,10-phenanthroline to activated Ir-complexes, which resulted in complete suppression of substrate exchange.[46] However, hydrogen exchange remains active as it was later demonstrated that the substrates on the Ir-complex can still be polarized[47]. In the present work, with bidentate pyruvate binding, we have a similar situation. At low temperatures of -8°C there is no observable pyruvate dissociation on the NMR timescale of several seconds,[22] however, the complex-bound pyruvate continues to be polarized. Considering that all evidence points to the need for substrate dissociation before hydrogen exchange,[37,48] we needed to introduce alternative possibilities for this process: partial pyruvate dissociation (**fig 1b**) and co-ligand elimination (**fig 1c**). To deduce which method is more probable, DFT simulations similar to the ones made for the prototypical Ir-complex with pyridine,[36] or more detailed exchange measurements are needed. The critical conclusion from this discussion for the present work is that the observed signals predominantly stem from catalyst-bound pyruvate.

Despite the excellent resolution with a full width at half maximum (FWHM) of 0.3 Hz, we could not identify any hydride-$^{13}$C $J$-coupling constants. Accordingly, we used $J_{HC}$ = 0.06 Hz value for the simulations, which is below the FWHM of the $^{13}$C lines of 1-$^{13}$C-pyruvate (see SI for further details). This value is two orders of magnitude below the one used before to simulate the spin evolution of this system (5 Hz in Ref [23]).

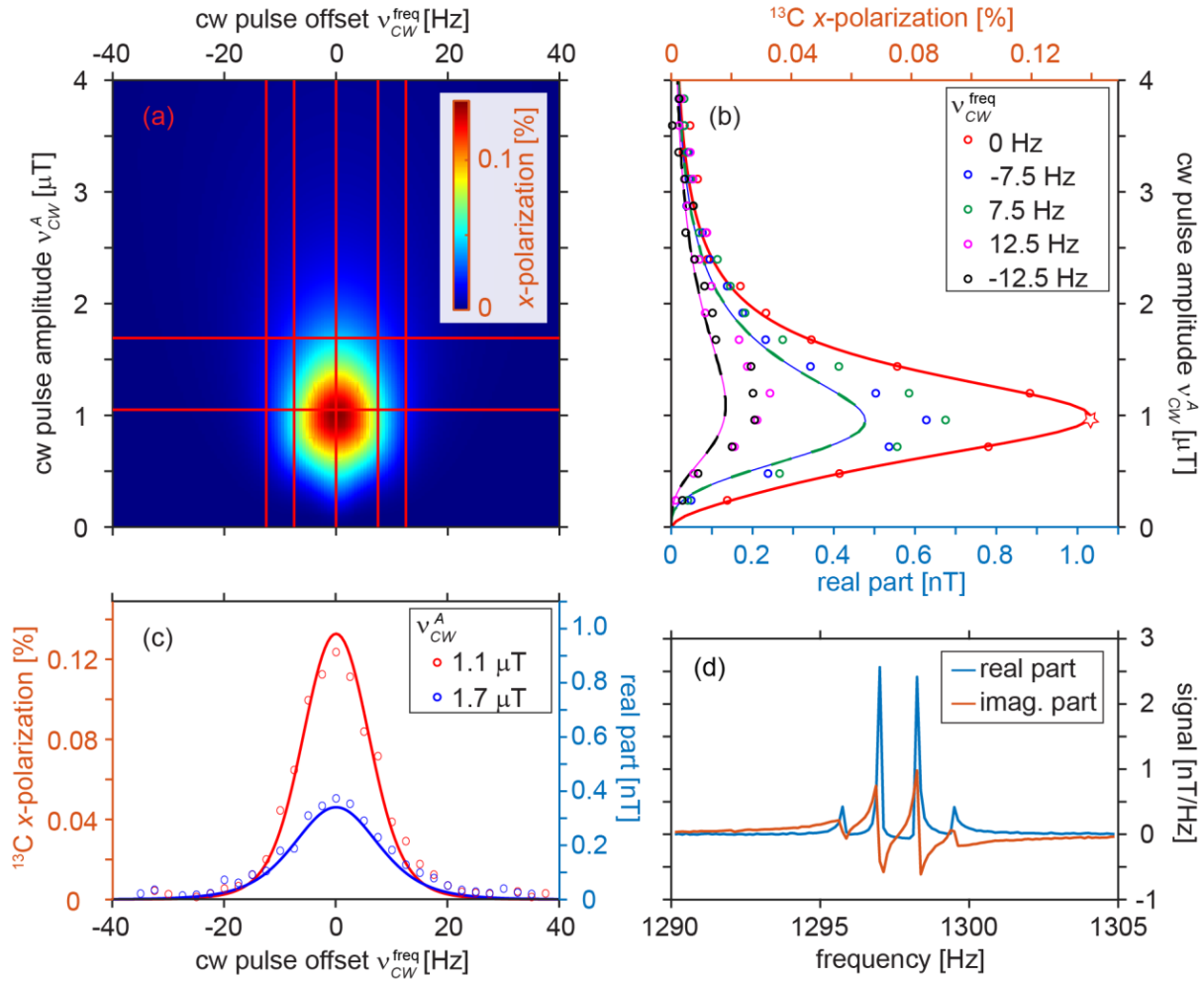

FIGURE 2. Simulated (a) and experimental with simulated (b, c, d) transverse (x-) polarization of 1-$^{13}$C-pyruvate by LIGHT-SABRE at $B_0 = 121$ µT as a function of the amplitude, $\nu_{CW}^{A}$, and frequency offset from the $^{13}$C resonance, $\Delta\nu_{CW}^{frq}$, of the CW pulse. Simulated x-polarization (a) with red lines indicating the position where experiments were conducted (b, c, circles). The highest polarization was observed at $B_1 \approx 1$ µT and on resonance (b, star). Corresponding hyperpolarized spectrum (d) showed the expected splitting because of the $^{13}$C-$^{1}$H J-coupling interactions. Note that the simulations were scaled on b and c to match the data point at $\Delta\nu_{CW}^{frq} = 0$. The parameters used in the simulation were $\tau_{Ir} = 31$ ms, $J_{HC} = 0.06$ Hz (one hydride to $^{13}$C, the other coupling is 0), $J_{HH} = -10.5$ Hz (hydride-hydride), $J_{C-H_3} = 1.2$ Hz ($^{13}$C to methyl protons). The spin system consisted of five protons and one carbon. Hyperpolarization time was 10 s, which is only a fraction of the full build-up time with a build-up time-constant of $T_{hyp} = 26$ s (see **fig.**

3). Also note that the $^{13}$C polarization was averaged across free and bound pyruvate, and the bound polarization was estimated to be ten times larger.

Theoretically, the creation of transverse *x*-polarization, when applying a $B_1$ field on resonance, along *x*-axis, can be rationalized by examining the following portion of the governing Hamiltonian as fully derived in the SI:

$$\begin{array}{cc} & \begin{array}{cc} |S_0\,X_-\rangle & |T_0\,X_+\rangle \end{array} \\ \begin{array}{c} |S_0\,X_-\rangle \\ |T_0\,X_+\rangle \end{array} & \begin{pmatrix} -J_{HH} - v_{CW}^A/2 & \Delta J_{CH}/4 \\ \Delta J_{CH}/4 & +v_{CW}^A/2 \end{pmatrix} \end{array}$$

Here $|S_0\rangle$ and $|T_0\rangle$ are states of the hydride protons with a longitudinal projection of the total spin of zero. $|X_-\rangle = \frac{|\alpha\rangle - |\beta\rangle}{\sqrt{2}}$ and $|X_+\rangle = \frac{|\alpha\rangle + |\beta\rangle}{\sqrt{2}}$ are antiparallel and parallel $^{13}$C states with respect to the CW field. $|\alpha\rangle$ and $|\beta\rangle$ are parallel and antiparallel $^{13}$C states with respect to the $B_0$ magnetic field.

Under the condition that the frequency offset, $\Delta v_{CW}^{frq}=0$, and the amplitude of CW is exactly on resonance with the *J*-coupling interaction, i.e. $v_{CW}^A = -J_{HH}$, the difference of the diagonal elements becomes zero and the off-diagonal element, $\Delta J_{CH}/4$ (the difference between the two hydride-$^{13}$C *J*-couplings), can efficiently couple $|S_0\,X_-\rangle$ to $|T_0\,X_+\rangle$ and drive spin alignment from the pH$_2$ into polarization along the applied $B_1$ field on the $^{13}$C nucleus.

To benchmark the new field-cycling-free LIGTH-SABRE experiment vs. the established SABRE-SHEATH method, we measured the build-up of hyperpolarization using both methods under otherwise identical conditions (**fig. 3**). SABRE-SHEATH seems to have a slightly faster build-up rate of (15.8 ± 2.9) s than LIGHT-SABRE with a build-up rate of (25.8 ± 2.1) s for $^{13}$C polarization. Remarkably, LIGHT-SABRE gives a hyperpolarization level that is comparable to that of SABRE-SHEATH. The achieved hyperpolarization level is 1.1 % when averaged over free and bound pyruvate. However, considering that the signal predominantly stems from bound pyruvate, which is at a ten-fold lower concentration, the hyperpolarization was estimated to be 11% for the bound pyruvate. The fact that both LIGHT-SABRE and SABRE-SHEATH on the same system allowed for close to identical signal enhancement indicates that both techniques are of the same polarization transfer efficiency for the pyruvate SABRE system.

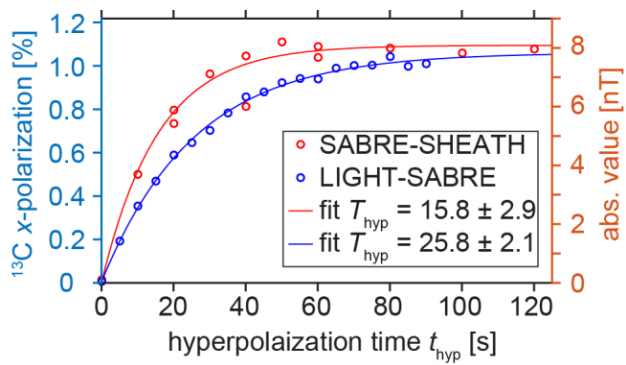

FIGURE 3. Polarization build-up in a SABRE-SHEATH (red circles) and in LIGHT-SABRE (blue circles) experiments at optimal conditions as a function of hyperpolarization time $t_{hyp}$ and fit (lines). The maximum polarization of $p \sim 1.1\%$ was achieved with SABRE-SHEATH and LIGHT-SABRE. $p = 1.1\%$ is averaged for free and catalyst-bound pyruvate. Catalyst-bound pyruvate polarization is estimated to be $p \approx 11\%$. The fitted constants for the mono exponential build-up were 15.8 s and 25.8 s, correspondingly. The polarization field for SABRE-SHEATH was ~0.36 µT. The CW parameters: $\Delta\nu_{CW}^{frq}=0$ and $\nu_{CW}^{A} \sim 11$ Hz, were used for LIGHT-SABRE.

Next, we explored the creation of z-polarization with LIGHT-SABRE (**fig 4**). The advantage of creating z-polarization is that the spins will not dephase as quickly once created, also because they are subject to $T_1$ decay and not to $T_{1\rho}$ effects, such that it may be easier to build up more magnetization simply by applying longer LIGHT-SABRE pulses. The observation of $^{13}C$ z-polarization requires a 90° pulse after CW. To avoid any contribution to the signal from transverse components, we also implemented a magnetic field gradient, which purposefully dephased all transverse spin orders generated with the LIGHT-SABRE before applying a 90º pulse to assess the longitudinal spin order only.

Two extrema at $\Delta\nu_{CW}^{frq} = \pm 6.3$ Hz were observed for longitudinal $^{13}$C magnetization both in experiments and simulations, which were found to match nicely (**fig. 4**). $^{13}$C $z$-Polarization on the order of 0.04 % was observed, which remained below 0.12% achieved for $x$-polarization with identical hyperpolarization time $t_{hyp}$. Note that in **fig. 2** and **4** the hyperpolarization time was 10 s, while the maximum polarization of 1.1% (averaged across free and bound, catalyst-bound $p \approx 11\%$) was reached at about 80 s (**fig. 3**).

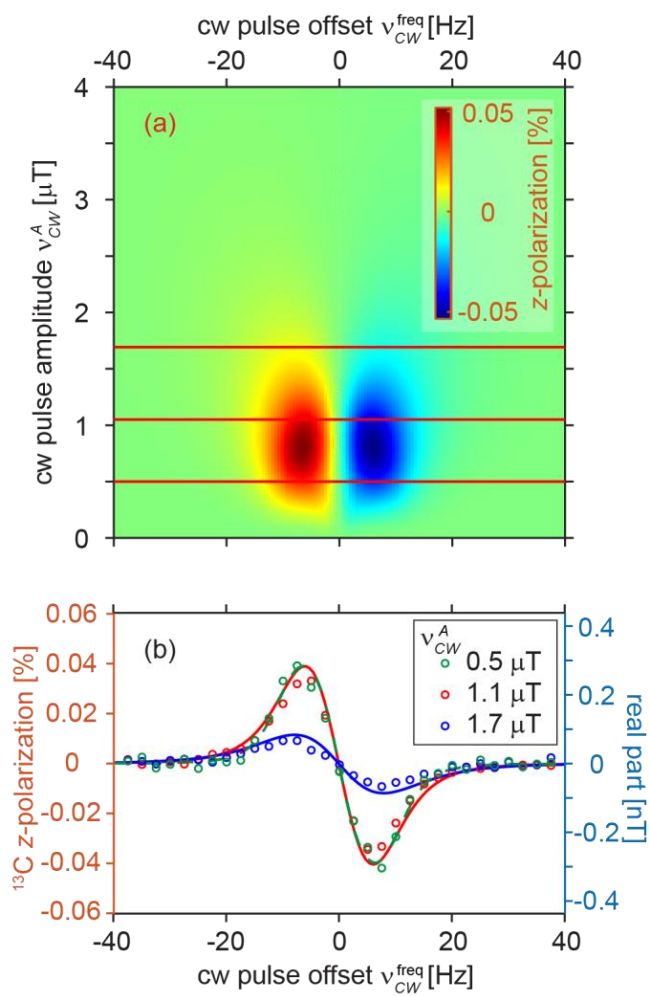

FIGURE 4. Simulated (a) and measured with simulated (b) longitudinal z-polarization of $^{13}$C as a function of $\Delta\nu_{CW}^{frq}$ and $\nu_{CW}^{A}$. The measurements were fit to simulations from (a) (red lines indicate the position). The maxima correspond to $\Delta\nu_{CW}^{frq} = \pm 6.3$ Hz and $\nu_{CW}^{A} \sim 11$ Hz. Hyperpolarization time was 10 s. $^{13}$C polarization was averaged across free and bound pyruvate; the bound polarization was estimated to be ten times larger.

Theoretically, the creation of z-polarization, when applying a $B_1$ field slightly off-resonance $\Delta\nu_{CW}^{frq}$, can be rationalized by examining the following portion of the governing Hamiltonian represented in a rotating frame of reference with a tilted axis for the $^{13}$C spin around the y-axis by an angle $\theta = \arctan\left(\frac{\nu_{CW}^{A}}{\Delta\nu_{CW}^{frq}}\right)$. Then, the effective field experienced by $^{13}$C is $\nu_{eff} = \sqrt{\left(\Delta\nu_{CW}^{frq}\right)^2 + \left(\nu_{CW}^{A}\right)^2}$. The basis for $^{13}$C in this tilted frame will be given by states: $|Z'_+\rangle = \cos(\theta/2)|\alpha\rangle + \sin(\theta/2)|\beta\rangle$ and $|Z'_-\rangle = \sin(\theta/2)|\alpha\rangle - \cos(\theta/2)|\beta\rangle$. The corresponding governing Hamiltonian block of relevance appears as follows (also fully explained in the SI):

$$\begin{array}{c} \phantom{|S_0Z'_-\rangle} \quad |S_0Z'_-\rangle \qquad\qquad |T_0Z'_+\rangle \\ \begin{array}{c} |S_0Z'_-\rangle \\ |T_0Z'_+\rangle \end{array} \left( \begin{array}{cc} -J_{HH} - \frac{\nu_{eff}}{2} & \Delta J_{CH}\sin(\theta)/4 \\ \Delta J_{CH}\sin(\theta)/4 & +\frac{\nu_{eff}}{2} \end{array} \right) \end{array}$$

Under the condition that $B_1$ is applied such that $\nu_{eff} = -J_{HH}$, the difference of the diagonal elements becomes zero, and the off-diagonal element can efficiently couple the states and drive polarization from the parahydrogen-derived singlet into *partial z-polarization*. As is fully detailed in the SI, the creation of z-polarization is most efficient when $\sin(2\theta)\times\sin(\theta)$ is maximized, which occurs at "the magic angle" of $\theta \cong 54.7°$ or for $\tan(\theta) = \sqrt{2}$. This theoretical finding is experimentally substantiated in **fig. 4b**, where the combination of offset $\Delta\nu_{CW}^{frq}$ and amplitude $\nu_{CW}^{A}$ at values close to the magic angle give the highest polarization levels.

Finally, to observe $^1$H-$^{13}$C zz-polarization, we excited only $^1$H with a 90º pulse that resulted in antiphase spectra. Alternatively, we also used the selective excitation of polarization using PASADENA (SEPP) sequence[49,50]. SEPP converts the antiphase spectral lines (**fig. 5**) into in-phase spectra, which can be used for $^1$H imaging, for example. As detailed in **fig. 5**, a good match between the experiment and simulation was obtained for the creation of zz-polarization.

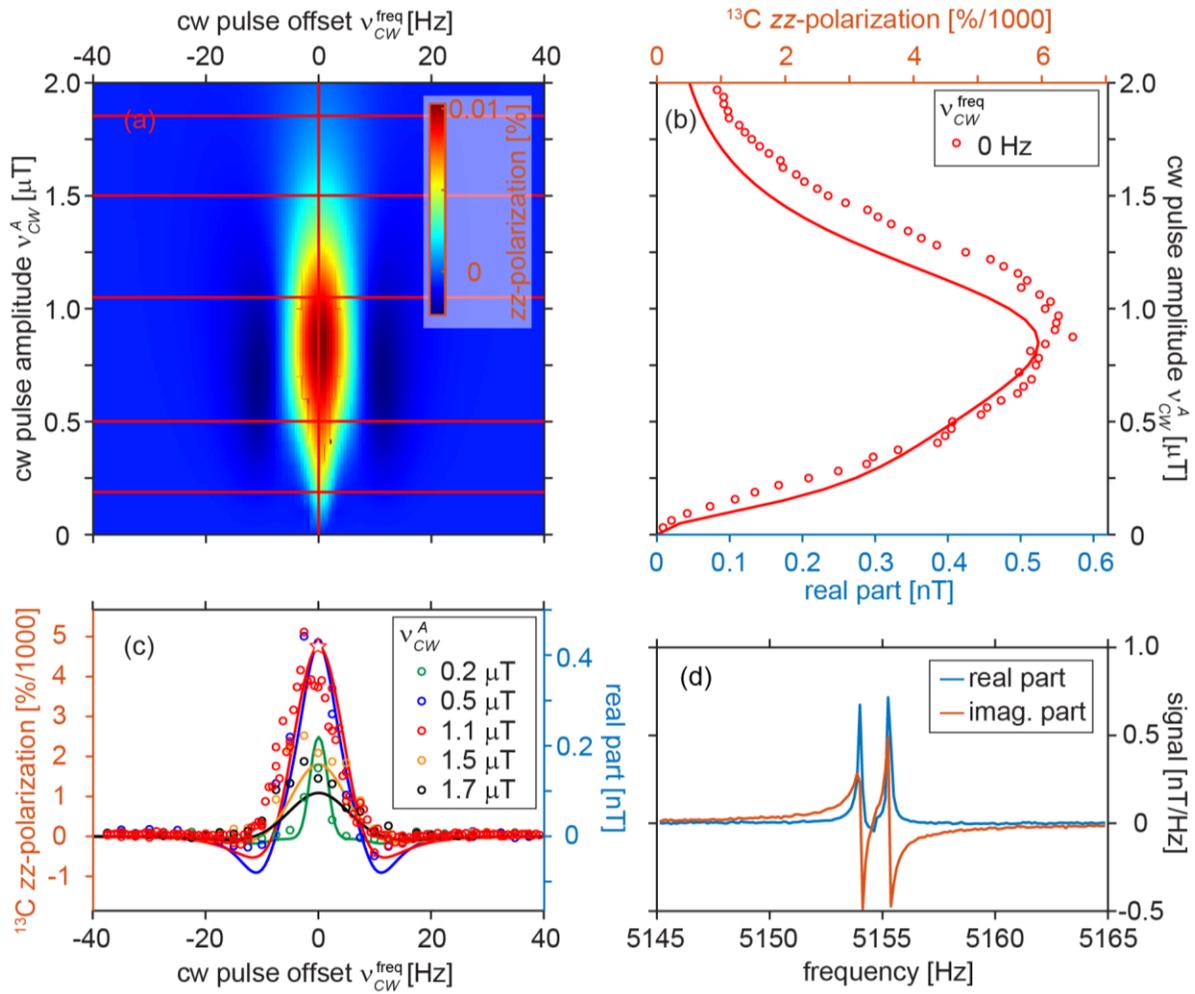

FIGURE 5: Simulated $^1$H-$^{13}$C zz-polarization as a function of $\Delta\nu_{CW}^{frq}$ and $\nu_{CW}^{A}$ (a) and experimentally measured with simulated zz-polarization as a function of the amplitude, $\nu_{CW}^{A}$, and frequency offset from $^{13}$C resonance, $\Delta\nu_{CW}^{frq}$, of the CW pulse (b, c). An exemplary $^1$H SEPP spectrum is shown in (d) for the case of maximum polarization (indicated

by the star in (c)). Polarization values were averaged over free and bound pyruvate; the bound polarization was estimated to be ten times larger.

To conclude, a new method for the continuous hyperpolarization of 1-$^{13}$C-pyruvate was demonstrated, and $^{13}$C polarization of more than 1 % was obtained after weak irradiation at the $^{13}$C Larmor frequency. Since pyruvate exchange is suppressed under the current experimental conditions of -8° C, the bound pyruvate hyperpolarization is estimated to be 11%. In this first proof-of-concept demonstration, the generation of *x*-polarization under on-resonance conditions was higher than the direct generation of *z*-magnetization under slight off-resonance conditions. For the generation of *x*- and *z*-polarization, we provided the theoretical underpinning by exploring the states that are coupled. Furthermore, $^{1}$H-$^{13}$C two-spin order was revealed. In our experimental setup, similar polarization levels were obtained with LIGHT-SABRE and control-SABRE-SHEATH experiments. Although the ultra-low field system has a superior magnetic field homogeneity, giving linewidths of below 0.3 Hz, we couldn't measure the $^{1}$H-$^{13}$C *J* coupling interaction, which is responsible for polarization transfer in SABRE-SHEATH and LIGHT-SABRE experiments. Furthermore, we discussed that pyruvate does not exchange into its free form at the employed low temperatures. However, the bound pyruvate is still polarized efficiently, implying that hydrogen still exchanges at a significant rate. With these insights, we proposed two mechanisms for the necessary H$_2$ exchange process: partial cleavage of pyruvate or axial co-ligand (DMSO) elimination. We suspect the partial pyruvate dissociation is the more likely explanation and intend to substantiate this claim with future theoretical and experimental evidence.

In future extrapolation to spin systems that may have additional coupled spins, e.g., other $^{1}$H, $^{2}$H or $^{31}$P, SABRE-SHEATH would transfer polarization to all of them,[14,51–53] while LIGHT-SABRE transfers polarization from pH$_2$ only to the irradiated spins, making the process more focused. We also observed this effect, simulating the effect of adding the methyl protons to the system. Addition of the methyl protons in the simulations reduced SABRE-SHEATH polarization more than LIGHT-SABRE polarization (**Table S2, SI**). Importantly, LIGHT-SABRE can avoid spin order transfer to relaxation sinks such as quadrupolar nuclei (e.g. $^2$H or $^{14}$N),[54,55] which often pose a challenge in SABRE-SHEATH.[56]

These considerations imply that for more complex spin systems, the LIGHT-SABRE technique can be expected to be more efficient than SABRE-SHEATH.

In our experiments conducted at $B_0$ = 121 µT, the Larmor frequency of $^{13}$C was about 1300 Hz. Thus, the irradiation frequency of the present demonstration lies in the audio frequency range. This is to the benefit of the LIGHT-SABRE approach because a simple audio source like a sound card can generate the necessary irradiation for polarization transfer without heating up the sample. In the case of dDNP, microwave sources, and microwave guides are necessary.

One of the limitations of LIGHT-SABRE and similar methods is that they are frequency-selective. Hence the efficiency is reduced when $B_0$ field homogeneity cannot be maintained. At ultralow magnetic fields, constant pH$_2$ bubbling did not deteriorate $B_0$ homogeneity.

In this work, low temperatures (-8°C) were used to allow for more efficient polarization transfer to bound pyruvate achieved at the reduced hydrogen exchange rates. A consequence of the low temperatures is almost complete suppression of pyruvate exchange, which, in future experiments, could also be compensated by using higher pH$_2$ pressures. In the current system, we only had atmospheric pH$_2$ pressure available, and more polarization may be expected at higher pressures and faster flow rates.[57] Because the *J*-coupling interaction of pyruvate with pH$_2$ is much smaller than $1/\tau_{Ir}$, the apparent hydrogen exchange rate, it appears that more studies in the direction of optimization of the catalyst and co-ligand (here DMSO) are necessary. At present, potentially effective strategies to harness the high degrees of bound pyruvate hyperpolarization are either temperature cycling as performed in the context of SABRE-SHEATH[22] or the addition of highly competitive ligands after hyperpolarization to displace the bound, hyperpolarized pyruvate from the catalyst.

Finally, we note that the used carben Ir-complex was primarily optimized for the polarization of pyridine-like molecules.[37,58] With novel optimized catalysts, the SABRE spin order transfer sequences developed might also be used to generate hyperpolarization at higher fields outside of the µT regime.

**EXPERIMENTAL AND COMPUTATIONAL DETAILS**

**Pulse sequences.** Since the sample degrades over time (**SI**, **fig. S1**) the performance of the system was monitored with a SABRE-SHEATH sequence (**fig. 6a**) every 5-10 minutes.

For a direct comparison between the methods SABRE-SHEATH (**fig. 6a**) and LIGHT-SABRE (**fig. 6b**) the hyperpolarization time $t_{hyp}$ was varied. After the SABRE-SHEATH phase, the use of a 90° pulse flipping $^1$H and $^{13}$C spins enabled simultaneous observation of longitudinal signals for $^1$H and $^{13}$C, whereas after the LIGHT-SABRE phase, no additional excitation was required for detection of the transverse magnetization.

To compare with the theoretical SABRE model, the LIGHT-SABRE sequence was repeated multiple times, varying frequency offset from $^{13}$C resonance $\Delta\nu_{CW}^{frq}$ and amplitude $\nu_{CW}^{A}$. Transverse $^{13}$C magnetization generated in LIGHT-SABRE was measured without additional excitation (**fig. 6c(i)** read-out scheme), while longitudinal magnetization was measured after crusher gradient dephasing remaining transverse magnetization from the SLIC pulse followed by a 90° pulse on $^1$H and $^{13}$C (**fig. 6c(ii)**).

The $^1$H-$^{13}$C longitudinal two-spin order was measured after crusher gradient using a $^1$H-$^{13}$C SEPP sequence (**fig. 6c(iii)** read-out scheme).

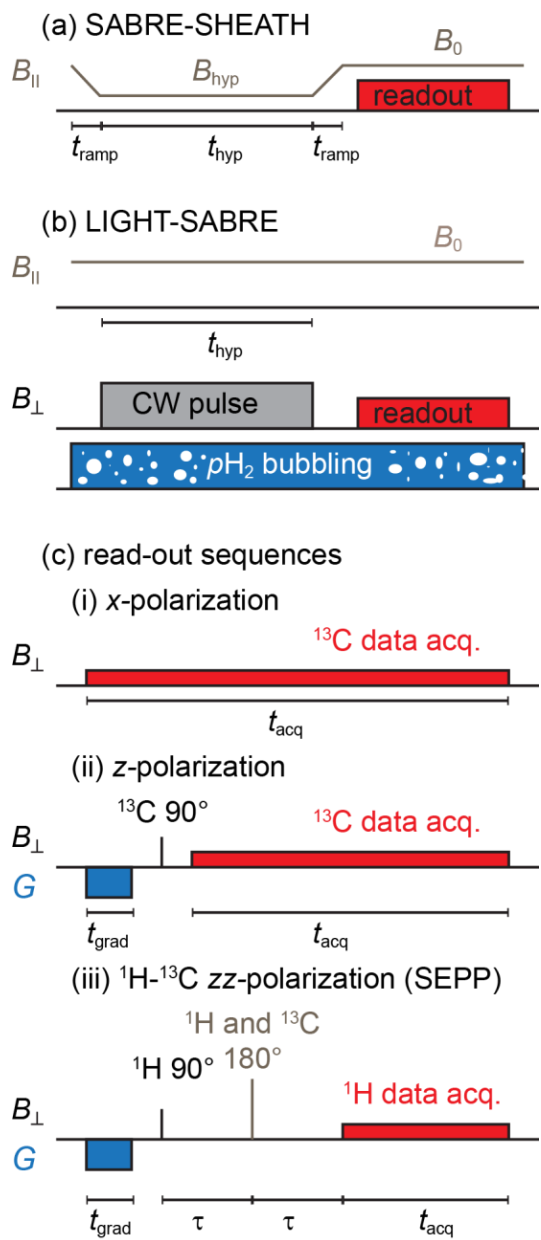

FIGURE 6: Schematic of a SABRE-SHEATH (a) and LIGHT-SABRE (b) sequences. The LIGHT-SABRE sequence can be combined with three read-out schemes (c, i-iii). For sole observation of transverse $^{13}$C magnetization no further pulses are needed (i). For observation of longitudinal magnetization of $^{13}$C and $^{1}$H, the transverse magnetization must be dephased via a crusher gradient before $^{1}$H, $^{13}$C 90° excitation pulse (ii). Two-spin order longitudinal $^{1}$H-$^{13}$C spin order was measured after a SEPP sequence (iii). Note that in SEPP there is a 90° pulse only on $^{1}$H, whereas the 180° pulse flips $^{1}$H and $^{13}$C spins.

**Sample.** To demonstrate the feasibility of LIGHT-SABRE on 1-$^{13}$C-pyruvate we prepared a sample consisting of 51 mM 1-$^{13}$C sodium pyruvate, 5.1 mM [Ir(COD)(IMes)Cl] SABRE pre-catalyst (IMes=1,3-bis(2,4,6-trimethylphenyl)imidazol-2-ylidene, COD =1,5-cyclooctadiene) and 18 mM dimethyl sulfoxide (DMSO) dissolved in methanol-H$_4$. Close to100% enrichment pH$_2$ was prepared using an in-house built liquid helium pH$_2$ generator[40].

**Spin simulation.** To simulate the SABRE experiment, we used the formalism developed for the linear exchange model.[59] Note the critical difference, we assume that only H$_2$ exchanges, while the substrate stays bound to Ir-complex. This can be expressed with the following chemical reaction

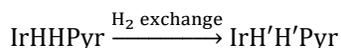

$$\text{IrHHPyr} \xrightarrow{\text{H}_2 \text{ exchange}} \text{IrH}'\text{H}'\text{Pyr}$$

where IrHHPyr states for the active SABRE complex and prime symbol ( ′) is used to distinguish hydrogens before and after exchange. For this chemical exchange, the following generalized Liouville-von Neuman equation has to be solved numerically:

$$\frac{d\hat{\rho}_{\text{IrHHPyr}}}{dt} = \left[\hat{\hat{L}}_{\text{IrHHPyr}} + \frac{1}{\tau_c}\left(\hat{\hat{D}}^{\text{H}_2^*}_{\text{IrPyr}} \widehat{\widehat{Tr}}^{\text{HH}}_{\text{IrHHPyr}} - \hat{\hat{1}}_{\text{IrHHPyr}}\right)\right]\hat{\rho}_{\text{IrHHPyr}}$$

Here $\hat{\rho}_{\text{IrHHPyr}}$ is the density matrix for the active Ir-complex with H$_2$ and pyruvate (Pyr), $\hat{\hat{L}}$ is corresponding Liouville superoperator, $\hat{\hat{1}}$ is a unitary superoperator, $\hat{\hat{D}}^{\text{H}_2^*}_{\text{IrPyr}}$ is the superoperator of the direct product that adds pH$_2$ to IrPyr and $\widehat{\widehat{Tr}}^{\text{HH}}_{\text{IrHHPyr}}$ is a superoperator that removes HH from IrHHPyr, resulting in a density matrix for IrPyr only, $1/\tau_{\text{Ir}}$ is the exchange rate of H$_2$, $\tau_{\text{Ir}}$ is the lifetime of the complex that was introduced in the text before.

The simulations were fitted to the experimental observation, varying $\tau_{\text{Ir}}$ and were scaled up to fit the experimental polarization values. Note that theoretical predictions result in an overestimation of the $^{13}$C polarization level. Predicted polarization levels were multiplied by a factor of 0.38 for $^{13}$C and 3.4 for $^1$H to fit the experimentally observed polarization. The system parameters are detailed in **Table S2, SI.**

In this simulation model, we completely neglected the transient complexes presented in **Fig. 1b-c**. Also, we assumed that pure pH$_2$ substitutes dihydrogens in Ir-complex after each exchange event. Both effects were simulated before in more detail.[45] However, because very little is known about these evolution steps of the Ir-complex, we used the most straightforward exchange scheme given above.

## ASSOCIATED CONTENT

The following files are available free of charge as Supporting Information: MOIN-spin library with all simulation scripts (.zip) and additional experimental results (.pdf).

## AUTHOR INFORMATION

## AUTHOR CONTRIBUTIONS

*ANP, KB, TT conceptualization, planning; KB, NK experiments, KB data evaluation and preparation of visuals, ANP simulations, ANP, TT analytical SLIC description, MP catalyst synthesis, ANP, KB, TT preparation of the first draft. All authors contributed to discussions and helped interpret the results, and gave approval to the final version of the manuscript.*

**Notes**

## ACKNOWLEDGMENT

ANP and JBH acknowledge funding from German Federal Ministry of Education and Research (BMBF) within the framework of the e:Med research and funding concept (01ZX1915C), DFG (PR 1868/3-1, HO-4604/2-2, HO-4604/3, GRK2154-2019, EXC2167, FOR5042, TRR287). MOIN CC was funded by a grant from the European Regional Development Fund (ERDF) and the Zukunftsprogramm Wirtschaft of Schleswig-Holstein (Project no. 122-09-053). ANP, KB, MP and RK acknowledge funding from DFG grant Nr. BU 2694/6-1. All authors thank the project DEAL for open-access publication. T. T. acknowledges funding from the National Institute of Biomedical Imaging and Bioengineering of the National Institutes of Health under Award No. NIH R01EB029829. The content is solely the responsibility of the authors and does not necessarily represent the official views of the National Institutes of Health. T. T. also acknowledges funding from the Goodnight foundation as well as the Alexander von Humboldt foundation.

REFERENCES

(1) Nelson, S. J.; Kurhanewicz, J.; Vigneron, D. B.; Larson, P. E. Z.; Harzstark, A. L.; Ferrone, M.; Criekinge, M. van; Chang, J. W.; Bok, R.; Park, I.; Reed, G.; Carvajal, L.; Small, E. J.; Munster, P.; Weinberg, V. K.; Ardenkjaer-Larsen, J. H.; Chen, A. P.; Hurd, R. E.; Odegardstuen, L.-I.; Robb, F. J.; Tropp, J.; Murray, J. A. Metabolic Imaging of Patients with Prostate Cancer Using Hyperpolarized [1-13C]Pyruvate. *Sci. Transl. Med.* **2013**, *5* (198), 198ra108-198ra108. https://doi.org/10.1126/scitranslmed.3006070.

(2) Cunningham, C. H.; Lau, J. Y. C.; Chen, A. P.; Geraghty, B. J.; Perks, W. J.; Roifman, I.; Wright, G. A.; Connelly, K. A. Hyperpolarized 13C Metabolic MRI of the Human Heart. *Circ. Res.* **2016**, *119* (11), 1177–1182.

(3) Gallagher, F. A.; Woitek, R.; McLean, M. A.; Gill, A. B.; Garcia, R. M.; Provenzano, E.; Riemer, F.; Kaggie, J.; Chhabra, A.; Ursprung, S.; Grist, J. T.; Daniels, C. J.; Zaccagna, F.; Laurent, M.-C.; Locke, M.; Hilborne, S.; Frary, A.; Torheim, T.; Boursnell, C.; Schiller, A.; Patterson, I.; Slough, R.; Carmo, B.; Kane, J.; Biggs, H.; Harrison, E.; Deen, S. S.; Patterson, A.; Lanz, T.; Kingsbury, Z.; Ross, M.; Basu, B.; Baird, R.; Lomas, D. J.; Sala, E.; Wason, J.; Rueda, O. M.; Chin, S.-F.; Wilkinson, I. B.; Graves, M. J.; Abraham, J. E.; Gilbert, F. J.; Caldas, C.; Brindle, K. M. Imaging Breast Cancer Using Hyperpolarized Carbon-13 MRI. *Proc. Natl. Acad. Sci. U.S.A.* **2020**, *117* (4), 2092–2098. https://doi.org/10.1073/pnas.1913841117.

(4) Ardenkjær-Larsen, J. H.; Fridlund, B.; Gram, A.; Hansson, G.; Hansson, L.; Lerche, M. H.; Servin, R.; Thaning, M.; Golman, K. Increase in Signal-to-Noise Ratio of >10,000 Times in Liquid-State NMR. *Proc. Natl. Acad. Sci. U.S.A.* **2003**, *100* (18), 10158–10163. https://doi.org/10.1073/pnas.1733835100.

(5) Capozzi, A.; Cheng, T.; Boero, G.; Roussel, C.; Comment, A. Thermal Annihilation of Photo-Induced Radicals Following Dynamic Nuclear Polarization to Produce Transportable Frozen Hyperpolarized 13C-Substrates. *Nat. Commun.* **2017**, *8*, 15757.

(6) Ceillier, M.; Cala, O.; Daraï, T. E.; Cousin, S. F.; Stern, Q.; Guibert, S.; Elliott, S. J.; Bornet, A.; Vuichoud, B.; Milani, J.; Pages, C.; Eshchenko, D.; Kempf, J. G.; Jose, C.; Lambert, S. A.; Jannin, S. An Automated System for Fast Transfer and Injection of Hyperpolarized Solutions. *J. Magn. Reson. Open* **2021**, 100017. https://doi.org/10.1016/j.jmro.2021.100017.

(7) Ferrari, A.; Peters, J.; Anikeeva, M.; Pravdivtsev, A.; Ellermann, F.; Them, K.; Will, O.; Peschke, E.; Yoshihara, H.; Jansen, O.; Hövener, J.-B. Performance and Reproducibility of 13C and 15N Hyperpolarization Using a Cryogen-Free DNP Polarizer. *Sci. Rep.* **2022**, *12* (1), 11694. https://doi.org/10.1038/s41598-022-15380-7.

(8) Bastiaansen, J. A.; Merritt, M. E.; Comment, A. Real Time Measurement of Myocardial Substrate Selection in Vivo Using Hyperpolarized 13C Magnetic Resonance. *J. Cardiovasc. Magn. Reson.* **2015**, *17* (1), O15. https://doi.org/10.1186/1532-429X-17-S1-O15.

(9) Sharma, G.; Wen, X.; Maptue, N. R.; Hever, T.; Malloy, C. R.; Sherry, A. D.; Khemtong, C. Co-Polarized [1-13C]Pyruvate and [1,3-13C2]Acetoacetate Provide a Simultaneous View of Cytosolic and Mitochondrial Redox in a Single Experiment. *ACS Sens.* **2021**, *6* (11), 3967–3977. https://doi.org/10.1021/acssensors.1c01225.

(10) Qin, H.; Tang, S.; Riselli, A. M.; Bok, R. A.; Santos, R. D.; Criekinge, M. van; Gordon, J. W.; Aggarwal, R.; Chen, R.; Goddard, G.; Zhang, C. T.; Chen, A.; Reed, G.; Ruscitto, D. M.; Slater, J.; Sriram, R.; Larson, P. E. Z.; Vigneron, D. B.; Kurhanewicz, J. Clinical Translation of Hyperpolarized 13C Pyruvate and Urea MRI for Simultaneous Metabolic and Perfusion Imaging. *Magn. Reson. Med.* **2021**, *87*, 138–149. https://doi.org/10.1002/mrm.28965.

(11) Hövener, J.-B.; Pravdivtsev, A. N.; Kidd, B.; Bowers, C. R.; Glöggler, S.; Kovtunov, K. V.; Plaumann, M.; Katz-Brull, R.; Buckenmaier, K.; Jerschow, A.; Reineri, F.; Theis, T.; Shchepin, R. V.; Wagner, S.; Bhattacharya, P.; Zacharias, N. M.; Chekmenev, E. Y. Parahydrogen-Based Hyperpolarization for Biomedicine. *Angew. Chem. Int. Ed.* **2018**, *57* (35), 11140–11162. https://doi.org/10.1002/anie.201711842.

(12) Kovtunov, K. V.; Pokochueva, E.; Salnikov, O.; Cousin, S.; Kurzbach, D.; Vuichoud, B.; Jannin, S.; Chekmenev, E.; Goodson, B.; Barskiy, D.; Koptyug, I. Hyperpolarized NMR: D-DNP, PHIP, and SABRE. *Chem. Asian J.* **2018**, *13* (15), 1857–1871. https://doi.org/10.1002/asia.201800551.

(13) Korchak, S.; Emondts, M.; Mamone, S.; Bluemich, B.; Glöggler, S. Production of Highly Concentrated and Hyperpolarized Metabolites within Seconds in High and Low Magnetic Fields. *Phys. Chem. Chem. Phys.* **2019**, *21* (41), 22849–22856. https://doi.org/10.1039/C9CP05227E.

(14) Pravdivtsev, A. N.; Brahms, A.; Ellermann, F.; Stamp, T.; Herges, R.; Hövener, J.-B. Parahydrogen-Induced Polarization and Spin Order Transfer in Ethyl Pyruvate at High Magnetic Fields. *Sci. Rep.* **2022**, *12* (1), 19361. https://doi.org/10.1038/s41598-022-22347-1.

(15) Carrera, C.; Cavallari, E.; Digilio, G.; Bondar, O.; Aime, S.; Reineri, F. ParaHydrogen Polarized Ethyl-[1-13C]Pyruvate in Water, a Key Substrate for Fostering the PHIP-SAH Approach to Metabolic Imaging. *ChemPhysChem* **2021**, *22*, 1042. https://doi.org/10.1002/cphc.202100062.

(16) Ding, Y.; Korchak, S.; Mamone, S.; Jagtap, A. P.; Stevanato, G.; Sternkopf, S.; Moll, D.; Schroeder, H.; Becker, S.; Fischer, A.; Gerhardt, E.; Outeiro, T. F.; Opazo, F.; Griesinger, C.; Glöggler, S. Rapidly Signal-Enhanced Metabolites for Atomic Scale Monitoring of Living Cells with Magnetic Resonance. *Chem. Methods* **2022**, *2*, e202200023. https://doi.org/10.1002/cmtd.202200023.

(17) Chukanov, N. V.; Salnikov, O. G.; Shchepin, R. V.; Kovtunov, K. V.; Koptyug, I. V.; Chekmenev, E. Y. Synthesis of Unsaturated Precursors for Parahydrogen-Induced Polarization and Molecular Imaging of 1-13C-Acetates and 1-13C-Pyruvates via Side Arm Hydrogenation. *ACS Omega* **2018**, *3* (6), 6673–6682. https://doi.org/10.1021/acsomega.8b00983.

(18) Brahms, A.; Pravdivtsev, A.; Stamp, T.; Ellermann, F.; Sönnichsen, F.; Hövener, J.-B.; Herges, R. Synthesis of 13C and 2H Labeled Vinyl Pyruvate and Hyperpolarization of Pyruvate. *Chem. Eur. J.* **2022**, *28*, e202201210. https://doi.org/10.1002/chem.202201210.

(19) Adams, R. W.; Aguilar, J. A.; Atkinson, K. D.; Cowley, M. J.; Elliott, P. I. P.; Duckett, S. B.; Green, G. G. R.; Khazal, I. G.; López-Serrano, J.; Williamson, D. C. Reversible Interactions with Para-Hydrogen Enhance NMR Sensitivity by Polarization Transfer. *Science* **2009**, *323* (5922), 1708–1711. https://doi.org/10.1126/science.1168877.

(20) Iali, W.; Roy, S. S.; Tickner, B. J.; Ahwal, F.; Kennerley, A. J.; Duckett, S. B. Hyperpolarising Pyruvate through Signal Amplification by Reversible Exchange (SABRE). *Angew. Chem. Int. Ed.* **2019**, *58* (30), 10271–10275. https://doi.org/10.1002/anie.201905483.

(21) Adelabu, I.; TomHon, P.; Kabir, M. S. H.; Nantogma, S.; Abdulmojeed, M.; Mandzhieva, I.; Ettedgui, J.; Swenson, R. E.; Krishna, M. C.; Goodson, B. M.; Theis, T.; Chekmenev, E. Y. Order-Unity 13C Nuclear Polarization of [1-13C]Pyruvate in Seconds and the Interplay of Water and SABRE Enhancement. *ChemPhysChem* **2022**, *23*, e202100839. https://doi.org/10.1002/cphc.202100839.

(22) TomHon, P.; Abdulmojeed, M.; Adelabu, I.; Nantogma, S.; Kabir, M. S. H.; Lehmkuhl, S.; Chekmenev, E. Y.; Theis, T. Temperature Cycling Enables Efficient 13C SABRE-SHEATH Hyperpolarization and Imaging of [1-13C]-Pyruvate. *J. Am. Chem. Soc.* **2022**, *144* (1), 282–287. https://doi.org/10.1021/jacs.1c09581.

(23) Nantogma, S.; Eriksson, S. L.; Adelabu, I.; Mandzhieva, I.; Browning, A.; TomHon, P.; Warren, W. S.; Theis, T.; Goodson, B. M.; Chekmenev, E. Y. Interplay of Near-Zero-Field Dephasing, Rephasing, and Relaxation Dynamics and [1-13C]Pyruvate Polarization Transfer Efficiency in Pulsed SABRE-SHEATH. *J. Phys. Chem. A* **2022**, *126* (48), 9114–9123. https://doi.org/10.1021/acs.jpca.2c07150.

(24) Pravdivtsev, A. N.; Yurkovskaya, A. V.; Vieth, H.-M.; Ivanov, K. L.; Kaptein, R. Level Anti-Crossings Are a Key Factor for Understanding Para-Hydrogen-Induced Hyperpolarization in SABRE Experiments. *ChemPhysChem* **2013**, *14* (14), 3327–3331. https://doi.org/10.1002/cphc.201300595.

(25) Pravdivtsev, A. N.; Yurkovskaya, A. V.; Zimmermann, H.; Vieth, H.-M.; Ivanov, K. L. Transfer of SABRE-Derived Hyperpolarization to Spin-1/2 Heteronuclei. *RSC Adv.* **2015**, *5* (78), 63615–63623. https://doi.org/10.1039/C5RA13808F.

(26) Truong, M. L.; Theis, T.; Coffey, A. M.; Shchepin, R. V.; Waddell, K. W.; Shi, F.; Goodson, B. M.; Warren, W. S.; Chekmenev, E. Y. 15N Hyperpolarization by Reversible Exchange Using SABRE-SHEATH. *J. Phys. Chem. C* **2015**, *119* (16), 8786–8797. https://doi.org/10.1021/acs.jpcc.5b01799.

(27) Theis, T.; Truong, M. L.; Coffey, A. M.; Shchepin, R. V.; Waddell, K. W.; Shi, F.; Goodson, B. M.; Warren, W. S.; Chekmenev, E. Y. Microtesla SABRE Enables 10% Nitrogen-15 Nuclear Spin Polarization. *J. Am. Chem. Soc.* **2015**, *137* (4), 1404–1407. https://doi.org/10.1021/ja512242d.

(28) Zhivonitko, V. V.; Skovpin, I. V.; Koptyug, I. V. Strong 31P Nuclear Spin Hyperpolarization Produced via Reversible Chemical Interaction with Parahydrogen. *Chem. Commun.* **2015**, *51* (13), 2506–2509. https://doi.org/10.1039/C4CC08115C.

(29) Pravdivtsev, A. N.; Yurkovskaya, A. V.; Zimmermann, H.; Vieth, H.-M.; Ivanov, K. L. Enhancing NMR of Insensitive Nuclei by Transfer of SABRE Spin Hyperpolarization. *Chem. Phys. Lett.* **2016**, *661* (Supplement C), 77–82. https://doi.org/10.1016/j.cplett.2016.08.037.

(30) Knecht, S.; Kiryutin, A. S.; Yurkovskaya, A. V.; Ivanov, K. L. Efficient Conversion of Anti-Phase Spin Order of Protons into 15N Magnetisation Using SLIC-SABRE. *Mol. Phys.* **2018**, *0* (0), 1–10. https://doi.org/10.1080/00268976.2018.1515999.

(31) Pravdivtsev, A. N.; Skovpin, I. V.; Svyatova, A. I.; Chukanov, N. V.; Kovtunova, L. M.; Bukhtiyarov, V. I.; Chekmenev, E. Y.; Kovtunov, K. V.; Koptyug, I. V.; Hovener, J.-B. Chemical Exchange Reaction Effect on Polarization Transfer Efficiency in SLIC-SABRE. *J. Phys. Chem. A* **2018**. https://doi.org/10.1021/acs.jpca.8b07163.

(32) Ariyasingha, N. M.; Lindale, J. R.; Eriksson, S. L.; Clark, G. P.; Theis, T.; Shchepin, R. V.; Chukanov, N. V.; Kovtunov, K. V.; Koptyug, I. V.; Warren, W. S.; Chekmenev, E. Y. Quasi-Resonance Fluorine-19 Signal Amplification by Reversible Exchange. *J. Phys. Chem. Lett.* **2019**, 4229–4236. https://doi.org/10.1021/acs.jpclett.9b01505.

(33) Theis, T.; Truong, M.; Coffey, A. M.; Chekmenev, E. Y.; Warren, W. S. LIGHT-SABRE Enables Efficient in-Magnet Catalytic Hyperpolarization. *J. Magn. Reson.* **2014**, *248*, 23–26. https://doi.org/10.1016/j.jmr.2014.09.005.

(34) Trepakova, A. I.; Skovpin, I. V.; Chukanov, N. V.; Salnikov, O. G.; Chekmenev, E. Y.; Pravdivtsev, A. N.; Hövener, J.-B.; Koptyug, I. V. Subsecond Three-Dimensional Nitrogen-15 Magnetic Resonance Imaging Facilitated by Parahydrogen-Based Hyperpolarization. *J. Phys. Chem. Lett.* **2022**, *13* (44), 10253–10260. https://doi.org/10.1021/acs.jpclett.2c02705.

(35) R. Lindale, J.; L. Eriksson, S.; S. Warren, W. Phase Coherent Excitation of SABRE Permits Simultaneous Hyperpolarization of Multiple Targets at High Magnetic Field. *Phys. Chem. Chem. Phys.* **2022**, *24* (12), 7214–7223. https://doi.org/10.1039/D1CP05962A.

(36) Lin, K.; TomHon, P.; Lehmkuhl, S.; Laasner, R.; Theis, T.; Blum, V. Density Functional Theory Study of Reaction Equilibria in Signal Amplification by Reversible Exchange. *ChemPhysChem* **2021**, *22* (19), 1947–1957. https://doi.org/10.1002/cphc.202100204.

(37) Cowley, M. J.; Adams, R. W.; Atkinson, K. D.; Cockett, M. C. R.; Duckett, S. B.; Green, G. G. R.; Lohman, J. A. B.; Kerssebaum, R.; Kilgour, D.; Mewis, R. E. Iridium N-Heterocyclic Carbene Complexes as Efficient Catalysts for Magnetization Transfer from Para-Hydrogen. *J. Am. Chem. Soc.* **2011**, *133* (16), 6134–6137. https://doi.org/10.1021/ja200299u.

(38) Buckenmaier, K.; Scheffler, K.; Plaumann, M.; Fehling, P.; Bernarding, J.; Rudolph, M.; Back, C.; Koelle, D.; Kleiner, R.; Hövener, J.-B.; Pravdivtsev, A. N. Multiple Quantum Coherences Hyperpolarized at Ultra-Low Fields. *ChemPhysChem* **2019**, *20* (21), 2823–2829. https://doi.org/10.1002/cphc.201900757.

(39) Pravdivtsev, A. N.; Kempf, N.; Plaumann, M.; Bernarding, J.; Scheffler, K.; Hövener, J.-B.; Buckenmaier, K. Coherent Evolution of Signal Amplification by Reversible Exchange in Two Alternating Fields (Alt-SABRE). *ChemPhysChem* **2021**, *22*, 2381. https://doi.org/10.1002/cphc.202100543.

(40) Buckenmaier, K.; Rudolph, M.; Fehling, P.; Steffen, T.; Back, C.; Bernard, R.; Pohmann, R.; Bernarding, J.; Kleiner, R.; Koelle, D.; Plaumann, M.; Scheffler, K. Mutual Benefit Achieved by Combining Ultralow-Field Magnetic Resonance and Hyperpolarizing Techniques. *Rev. Sci. Instrum.* **2018**, *89* (12), 125103. https://doi.org/10.1063/1.5043369.

(41) DeVience, S. J.; Walsworth, R. L.; Rosen, M. S. Preparation of Nuclear Spin Singlet States Using Spin-Lock Induced Crossing. *Phys. Rev. Lett.* **2013**, *111* (17), 173002. https://doi.org/10.1103/PhysRevLett.111.173002.

(42) Svyatova, A.; Skovpin, I. V.; Chukanov, N. V.; Kovtunov, K. V.; Chekmenev, E. Y.; Pravdivtsev, A. N.; Hövener, J.-B.; Koptyug, I. V. 15N MRI of SLIC-SABRE Hyperpolarized 15N-Labelled Pyridine and Nicotinamide. *Chem. Eur. J.* **2019**, *25* (36), 8465–8470. https://doi.org/10.1002/chem.201900430.

(43) DeVience, S. J.; Walsworth, R. L.; Rosen, M. S. Dependence of Nuclear Spin Singlet Lifetimes on RF Spin-Locking Power. *Journal of Magnetic Resonance* **2012**, *218* (Supplement C), 5–10. https://doi.org/10.1016/j.jmr.2012.03.016.

(44) Pravdivtsev, A. N.; Yurkovskaya, A. V.; Lukzen, N. N.; Ivanov, K. L.; Vieth, H.-M. Highly Efficient Polarization of Spin-1/2 Insensitive NMR Nuclei by Adiabatic Passage through Level Anticrossings. *J. Phys. Chem. Lett.* **2014**, *5* (19), 3421–3426. https://doi.org/10.1021/jz501754j.

(45) Pravdivtsev, A. N.; Hövener, J.-B. Simulating Non-Linear Chemical and Physical (CAP) Dynamics of Signal Amplification By Reversible Exchange (SABRE). *Chem. Eur. J.* **2019**, *25* (32), 7659–7668. https://doi.org/10.1002/chem.201806133.
(46) Mewis, R. E.; Fekete, M.; Green, G. G. R.; Whitwood, A. C.; Duckett, S. B. Deactivation of Signal Amplification by Reversible Exchange Catalysis, Progress towards in Vivo Application. *Chem. Commun.* **2015**, *51* (48), 9857–9859. https://doi.org/10.1039/C5CC01896J.
(47) Pravdivtsev, A. N. SABRE Hyperpolarization of Bipyridine Stabilized Ir-Complex at High, Low and Ultralow Magnetic Fields. *Z. Phys. Chem.* **2016**, *231* (3), 497–511. https://doi.org/10.1515/zpch-2016-0810.
(48) Barskiy, D. A.; Pravdivtsev, A. N.; Ivanov, K. L.; Kovtunov, K. V.; Koptyug, I. V. A Simple Analytical Model for Signal Amplification by Reversible Exchange (SABRE) Process. *Phys. Chem. Chem. Phys.* **2016**, *18* (1), 89–93. https://doi.org/10.1039/C5CP05134G.
(49) Sengstschmid, H.; Freeman, R.; Barkemeyer, J.; Bargon, J. A New Excitation Sequence to Observe the PASADENA Effect. *J. Magn. Reson. A* **1996**, *120* (2), 249–257. https://doi.org/10.1006/jmra.1996.0121.
(50) Pravdivtsev, A.; Hövener, J.-B.; Schmidt, A. B. Frequency-Selective Manipulations of Spins for Effective and Robust Transfer of Spin Order from Parahydrogen to Heteronuclei in Weakly-Coupled Spin Systems. *ChemPhysChem* **2022**, *23*, e202100721. https://doi.org/10.1002/cphc.202100721.
(51) Marshall, A.; Salhov, A.; Gierse, M.; Müller, C.; Keim, M.; Lucas, S.; Parker, A.; Scheuer, J.; Vassiliou, C.; Neumann, P.; Jelezko, F.; Retzker, A.; Blanchard, J. W.; Schwartz, I.; Knecht, S. *Radio-Frequency Sweeps at {\mu}T Fields for Parahydrogen-Induced Polarization of Biomolecules*; arXiv:2205.15709; arXiv, 2022. https://doi.org/10.48550/arXiv.2205.15709.
(52) Cavallari, E.; Carrera, C.; Boi, T.; Aime, S.; Reineri, F. Effects of Magnetic Field Cycle on the Polarization Transfer from Parahydrogen to Heteronuclei through Long-Range J-Couplings. *J. Phys. Chem. B* **2015**, *119* (31), 10035–10041. https://doi.org/10.1021/acs.jpcb.5b06222.
(53) Dagys, L.; Bengs, C.; Moustafa, G. A. I.; Levitt, M. H. Deuteron-Decoupled Singlet NMR in Low Magnetic Fields: Application to the Hyperpolarization of Succinic Acid**. *ChemPhysChem* **2022**, e202200274. https://doi.org/10.1002/cphc.202200274.
(54) Barskiy, D. A.; Shchepin, R. V.; Tanner, C. P. N.; Colell, J. F. P.; Goodson, B. M.; Theis, T.; Warren, W. S.; Chekmenev, E. Y. The Absence of Quadrupolar Nuclei Facilitates Efficient 13C Hyperpolarization via Reversible Exchange with Parahydrogen. *ChemPhysChem* **2017**, *18* (12), 1493–1498. https://doi.org/10.1002/cphc.201700416.
(55) R. Birchall, J.; H. Kabir, M. S.; G. Salnikov, O.; V. Chukanov, N.; Svyatova, A.; V. Kovtunov, K.; V. Koptyug, I.; G. Gelovani, J.; M. Goodson, B.; Pham, W.; Y. Chekmenev, E. Quantifying the Effects of Quadrupolar Sinks via 15 N Relaxation Dynamics in Metronidazoles Hyperpolarized via SABRE-SHEATH. *Chemical Communications* **2020**. https://doi.org/10.1039/D0CC03994B.
(56) Korchak, S. E.; Ivanov, K. L.; Pravdivtsev, A. N.; Yurkovskaya, A. V.; Kaptein, R.; Vieth, H.-M. High Resolution NMR Study of T1 Magnetic Relaxation Dispersion. III. Influence of Spin 1/2 Hetero-Nuclei on Spin Relaxation and Polarization Transfer among Strongly Coupled Protons. *J. Chem. Phys.* **2012**, *137* (9), 094503. https://doi.org/10.1063/1.4746780.
(57) Colell, J. F. P.; Logan, A. W. J.; Zhou, Z.; Shchepin, R. V.; Barskiy, D. A.; Ortiz, G. X.; Wang, Q.; Malcolmson, S. J.; Chekmenev, E. Y.; Warren, W. S.; Theis, T. Generalizing, Extending, and Maximizing Nitrogen-15 Hyperpolarization Induced by Parahydrogen in Reversible Exchange. *J. Phys. Chem. C* **2017**, *121* (12), 6626–6634. https://doi.org/10.1021/acs.jpcc.6b12097.
(58) Rayner, P. J.; Norcott, P.; Appleby, K. M.; Iali, W.; John, R. O.; Hart, S. J.; Whitwood, A. C.; Duckett, S. B. Fine-Tuning the Efficiency of Para-Hydrogen-Induced Hyperpolarization by Rational N-Heterocyclic Carbene Design. *Nat. Commun.* **2018**, *9* (1), 4251. https://doi.org/10.1038/s41467-018-06766-1.
(59) Knecht, S.; Pravdivtsev, A. N.; Hövener, J.-B.; Yurkovskaya, A. V.; Ivanov, K. L. Quantitative Description of the SABRE Process: Rigorous Consideration of Spin Dynamics and Chemical Exchange. *RSC Adv.* **2016**, *6* (29), 24470–24477. https://doi.org/10.1039/C5RA28059A.